\documentclass[onecolumn]{aa} % for a paper on 1 column
\usepackage{graphicx}
\usepackage{txfonts}
\usepackage{longtable}
\begin{document}
   \title{Superluminal Motions of AGNs and GRBs at Multiple-Frequencies}

  % \subtitle{I. Overviewing the $\kappa$-mechanism}

   \author{Zhi-Bin Zhang
          \inst{1, 2}
   %      \and
%         Zhi-Yang Zhao\inst{1}
         %\fnmsep\thanks{Just to show the usage
 %         of the elements in the author field}
          }

   \institute{Department of Physics, College of Physics, Guizhou University, Guiyang, Guizhou 550025, China.\\
              \email{zbzhang@gzu.edu.cn}
         \and
             College of Physics and Engineering, Qufu Normal University, Qufu 273165, China.\\
          %   \email{c.ptolemy@hipparch.uheaven.space}
            % \thanks{The university of heaven temporarily does not
%                     accept e-mails}
             }

   %\date{Received February ??, 2010; accepted ?? }

% \abstract{}{}{}{}{}
% 5 {} token are mandatory

  \abstract
  % context heading (optional)
  % {} leave it empty if necessary
   {Superluminal motion has been found to occur in many kinds of celestial bodies
   with the relativistic ejecta, especially for some active galactic nuclei (AGNs), gamma-ray bursts (GRBs) and micro-quasars, although
   these objects are largely different in their lifetimes, sizes and features.
   }
  % aims heading (mandatory)
   {To investigate the discrepancy in above inhomogeneous objects, we have compared the apparent
   superluminal motion of GRBs with AGNs, both of them being cosmological outbursts. In addition,
   properties of the superluminal motion between different
   sub-classes of AGNs (e.g., Radio galaxies, BL Lac objects, and Quasars) are also compared in detail.Particularly, we focus on two low luminosity GRBs, namely 060218 and 170817A, that are ultra-long and short bursts associated with supernova and gravitational wave respectively.}
  % methods heading (mandatory)
   {For these, statistical methods including linear regression analysis, linear correlation and K-S test have been used.
  Antitheses and theoretical analysis of models are also adopted to contrast between AGNs and
   GRBs, as well as their individual subgroups at aspect of theory and observation. }
  % results heading (mandatory)
   {The apparent transverse velocity ($\beta_{app}$) of Swift and pre-Swift GRBs are tightly correlated with the
   Doppler factor $\delta$ as $\beta_{app}\propto\delta^{0.5}$, while all AGNs distribute around the line of
   $\beta_{app}= \delta$ behaving a weak correlation of
    $\beta_{app}$ with $\delta$ . In contrast, $\beta_{app}$ is positively correlated with Lorenz factor
    ($\gamma$) or $\gamma^2(1+z)^{-1}$, not for GRBs but for AGNs.
   $\beta_{app}$ and $1+z$ are independent for neither GRBs nor AGNs, but apparently exhibit a positive correlation of $\beta_{app}$ with $1+z$ from AGNs to GRBs,
  showing a cosmological effect of evolution with redshift. I also found that GRB 170817A and GRB 060218 are outliers of the above correlations.
  }
  % conclusions heading (optional), leave it empty if necessary
   {Superluminal properties of GRBs are significantly different from those of AGNs. However,
   there are no obvious differences between radio galaxies, BL Lac objects, and quasars in terms of their
    superluminal motion. In despite of radiation mechanism, beaming effect and cosmological expanding would
    play the same roles on the superluminal motion for AGNs and GRBs. }


   \maketitle
%
%________________________________________________________________

\section{Introduction}

The apparent superluminal phenomenon was first predicted by Rees
(1966) that the transverse velocity of an object moving
(ultra-)relativistically in some special directions may appear to
exceed the speed of light. At first glance such a motion is quite
counter-intuitive, whereas it does not violate the special
relativity. It is essentially a geometric effect or a light
travel-time effect in the frame of standard model (Chodorowski
2005).

This motion is initially confirmed and discovered by using new
technique of spectroscopy called as very long baseline
interferometry (VLBI) for radio galaxies and quasars (Whitney et al.
1971, Cohen et al. 1971). So far, it has been known that the
superluminal motion is not unique to quasars and radio galaxies, but
also to other sources including micro-quasars and BL Lac objects
(e.g. Mirabel \& Rodr\'{\i}guez 1994, Fan, Xie \& Wen 1996, Jorstad
et al. 2001, Kellermann et al. 2004) and so on. This may be because
jets or jet-like outflows are common among various kinds of
astrophysical phenomena with different scales, such as sun,
proto-stellar systems, isolated neutron stars, neutron star and
black hole binaries, supermassive black holes (or AGNs) and GRBs
(Zhang 2007). Once the apparent velocity is measurable, one can have
opportunity to learn the geometry and the underlying physics on the
formation, ejection and acceleration of jets (Ghisellini \& Celotti
A 2002).

%%%%%%%%%%%%%%%%%
AGNs are generally thought to originate from supermassive black
holes located in their centers. Classification is a very important
strategy to comprehend their basic physics. They can be classified
into several groups or subgroups by means of some diagnostic
diagrams, e.g. Seyfert I and II galaxies, Quasars, Blazars (BL lac
Objects and OVVs), radio galaxies, and LINER. On the whole, Blazars,
radio galaxies and part of Quasars are radio-loud, the remainder is
radio-quiet. The radio-loud AGNs typically exhibit strong radio
radiation, variable, jetted outflows, and X-ray emission and UV
excess. This is surprisingly similar to what have been observed in
GRBs. Moreover, these radio-loud sources together with GRBs have a
common property that the ejecta from their cores move very close to
the speed of light outwards and both of them are similarly
cosmological outbursts. Till now, only two GRBs, e.g. 030329 and 170817A, are promised to have
the detectably superluminal phenomenon in radio bands (Dado, Dar \& De Rujula 2004, 2018; Taylor et al. 2004).
Although the superluminal motions of GRBs in high energy afterglows even prompt $\gamma$-rays
are invisible currently, the superluminal effect for some GRBs would do exist in theory, as long
as the synergetic conditions such as viewing angles and boosting lorentz factors of outflows were satisfied (see Fig. 1).
This motivates us to give a comparative
study between GRBs and radio-loud AGNs on the superluminal motion.
Besides, GRBs in the pre-Swift and Swift eras have also been
compared to check the dependency of the apparent transverse velocity
on different instruments.
 \begin{figure}[h]
   \centering
 \includegraphics[width=12cm]{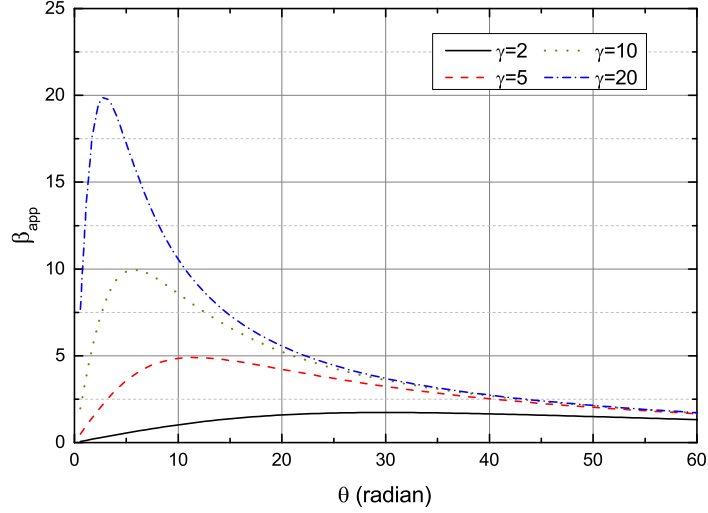}
      \caption{The apparent transverse velocity $\beta_{app}$ vs. the viewing angle $\theta$. Note that every $\beta_{app}>1$ corresponds a possible maximum angle $\theta_{max}=\arccos \beta$.}
      \label{Bapp-theta}
   \end{figure}

\section{Superluminal motion}

According to the relative geometry of ejecta to the observer, the
apparent transverse velocity $\beta_{app}$ is generally defined by
 \begin{equation}
     \beta_{app}=\frac{\beta\sin\theta}{1-\beta\cos\theta}
 \end{equation}
in an unit of speed of light $c$, of which $\beta$ represents the
real speed of ejecta and $\theta$ is the viewing angle between the
line of sight and the bulk velocity outwards. Note that the $\beta$
is also in the unit of $c$ and associated with Lorentz factor
$\gamma$ by $\beta=(1-\gamma^{-2})^{0.5}$. For a given $\gamma$, one
can make a plot of $\beta_{app}$ versus $\gamma$, where shows a
single peak profile at $\theta\approx1/\gamma$ (Cohen, Lister \&
Vermeulen 2004, Kellermann et al. 2007).

In addition to $\beta_{app}$, Doppler factor $\delta$ is also
connected with $\theta$ and $\gamma$ as
\begin{equation}
\delta=\frac{1}{\gamma(1-\beta\cos\theta)}\,,
\end{equation}

In Eqs. (1) and (2), four independent variables, i.e. $\beta_{app}$,
$\beta$, $\theta$ and $\delta$), are involved. Provided two of them
have been known in advance, the rest two should be deduced
immediately. For the radio observations to AGNs with measured
redshift ($z$), $\beta_{app}$ is in general measurable and
determined by
\begin{equation}
\beta_{app}=\frac{\mu D_A}{c}\,,
\end{equation}
where $\mu$ is the observed proper motion in an unit of mas/yr,
$D_A$ denotes the angular diameter distance of radio source. In
practice, the physical luminosity distance, $D_L$ in the flux
expression of $f=L/(4\pi D_L)$, is usually used for cosmological
studies. $D_A$ and $D_L$ are tied with each other as
\begin{equation}
D_L=(1+z)^2D_A\,,
\end{equation}
Hence, $\beta_{app}$ can then be newly expressed as
follows
\begin{equation}
\beta_{app}=\frac{\mu D_L}{c(1+z)^2}\,,
\end{equation}
The Doppler factor $\delta$ (Ghisellini et al. 1993) is estimated by
\begin{equation}
\delta=f(\alpha)S_m[\frac{ln(\nu_b/\nu_m)}{S_x\phi_{d}^{6+4\alpha}\nu_x^{\alpha}\nu_m^{5+3\alpha}}]^{1/(4+2\alpha)}\cdot(1+z)\,;
\end{equation}
where $f(\alpha)\simeq0.08\alpha+0.14$ is an experiential function
decided by an index $\alpha$ of the enhancement
($\propto\delta^{\alpha}$) of the observed flux due to beaming
effect, $S_m$ is the maximum flux at the frequency $\nu_m$ in GHz.
$S_x$ like $S_m$ in the unit of Jansky (Jy) represents the flux
within the range of any other energy band $\nu_x$, other than the
radio. $\nu_b$ and $\phi_d$ are the synchrotron high energy cutoff
and the size of core, respectively. As for AGNs, if only the
redshift can be gotten from the radio spectra one always calculate
$\beta_{app}$ and $\delta$, and then $\beta$ and $\theta$.

In principle, the superluminal phenomenon can be seen as long as the
viewing angle $\theta$ is suitably small enough in comparison with
the opening angle $\sim1/\gamma$ of outflows due to the beaming
action. For a given source, its opening angle is generally knowable
in a direct or indirect way, which makes Eq. (1) degenerate into
three scenarios in the following three cases of $\theta$ as:
 \begin{enumerate}
 \item In the case of $\theta\ll\gamma^{-1}$, the observer is enveloped within the cone of jets. Some
simplified forms of $\sin\theta\sim\theta$ and $\cos\theta\sim1$
lead Eq.(1) to be rewritten as
\begin{equation}
\beta_{app}\simeq2\gamma^2\theta
\end{equation}
 \item In the case of $\theta\sim\gamma^{-1}$, one can just see the edge of outflows under assumption
 that jets do not behave a significant expanding sideways. Relations of $\cos\theta\sim1-\gamma^{-2}/2$, $\sin\theta\sim\gamma^{-1}$ and $\beta=1-\gamma^{-2}/2$
 are derived. In this way, $\beta_{app}$ will be reduced and reach
 its maximum as
\begin{equation}
\beta_{app}\simeq\gamma\simeq\delta\simeq\csc\theta
\end{equation}
 \item In the case of $\theta\gg\gamma^{-1}$, the observer is located out of the cone of jets, which means $\sin\theta\sim\theta$, $\cos\theta\sim1-\theta^2/2$ and
 $\beta\sim1$, thus
\begin{equation}
\beta_{app}\simeq2/\theta
\end{equation}
\end{enumerate}
%__________________________________________________________

\section{Samples}

In view of above considerations, 8 radio galaxies, 15 BL Lac
objects, and 46 quasars with superluminal observations by VLBI have
been chosen from literatures (Hong et al. 1995, Jorstad et al. 2001,
Kellermann et al. 2004), to constitute three independent AGN samples
table 1.

For GRBs, present observations of radio afterglows hiding
superluminal motion are very rare. Till now, ONLY GRB 030329 and GRB
980425 are suggested to be the potential candidates (e.g. Dar \& De
Rujula 2000; Dado,  Dar \& De Rujula 2004). The reason is that the
angular resolution of current antennae is still too lower to measure
the proper motion of GRB afterglows, except for GRB 030329 ( Taylor
et al. 2004). It will be more difficult to describe the superluminal
motion in the phase of prompt emissions due to lack of much higher
resolution. This causes Eq.(3) to be impossible for deriving
$\beta_{app}$. Alternatively, Eq. (1) can be used to investigate the
superluminal motion during the prompt $\gamma$-ray emissions if only
the viewing angle $\theta$ can be conveniently measured provided the
Lorentz factors of GRBs are roughly estimated by
\begin{equation}
\gamma\approx240\times E_{iso, 52}^{1/8}n_{1}^{-1/8}(T_{90}/10
s)^{-3/8}\,,
\end{equation}
as suggested by Rees \& M\'{e}sz\'{a}ros (1992). Here, it has been
assumed that the jet opening angle $\theta_j$ is equivalent to the
viewing angle $\theta$ since the jetted outflow from us is much
farther than that from its central engine, which will give the upper
limits to estimate the apparent velocities of GRBs. As seen in Fig. 1,
larger lorentz factors together with wider viewing angles will result
in an enhancement of the superluminal motion.
Based on the consideration, the apparent superluminal motion of GRBs in their
prompt phase had been studied. For the sake of comparison, 27 Swift
and 37 pre-Swift long GRBs with known redshift, duration time and
jet break time are taken from Zhang, Zhao \& Zhang (2011). In addition, I pay particular attention
to GRB 170817A whose redshift, lorentz factor and viewing angle are estimated as
$z=0.0098$ (Levan et al. 2017), $\gamma\simeq13$ (Zou et al. 2018) and $30^{\circ}\geq\theta\geq23^{\circ}$ (Haggard et al. 2017; Ioka \& Nakamura 2017), respectively. Using these measured values, one can easily obtain the apparent transverse
velocity of GRB 170817A to be $\beta_{app}=4.2^{+0.5}_{-0.4}$ and the Doppler factor as $\delta=0.74^{+0.21}_{-0.18}$.

\onltab{1}{
\begin{table*}
\caption{Samples of Radio-Loud AGNs (B: Blazars; G: radio Galaxies; Q: Quassars)}             % title of Table
\label{tab1}      % is used to refer this table in the text
\centering                          % used for centering table
\begin{tabular}{cccccccc}        % centered columns (4 columns)
\hline\hline                 % inserts double horizontal lines
IAU & type &  $z$ & $\beta_{app}$& $\theta$ & $\delta$ & $\gamma$ & $\beta$  \\   % table heading
\hline                    % inserts single horizontal line
1219+285 &  B  & 0.102 &  2 &  36.41  & 1.56  &  2.38  & 0.90745\\
0219+428 &  B  & 0.444 & 14.98 &  5.8   & 1.99   & 12.34& 0.99671\\
0235+164 &  B  & 0.94  & 7.1 & 2.57 &   16.32 & 9.74 & 0.99472\\
0454+844 &  B  & 0.3   &1.6  & 8.49 &   4.46  & 2.63 & 0.92489  \\
0716+714 &  B  & 0.3 & 2.3 &  22.33 &2.63 & 2.51 & 0.91721  \\
0735+178 &  B  & 0.424 & 5.84 &6.39 &3.17 &  7.27& 0.99049  \\
0851+202 &  B & 0.306 &3.2 & 3.16 &  10.33 &5.71 & 0.98455  \\
0954+658 &  B  & 0.367& 5.7 & 8.61  & 6.62 & 5.84 &0.98523 \\
1101+384 &  B & 0.031 & 1.9 & 20.2  & 2.78 & 2.22 &  0.8928 \\
1308+326 &  B &0.996  & 20.7& 4.74 &  8.45 & 29.65& 0.99943  \\
1749+701 &  B &  0.77 & 6.12 & 17.76 &1.32& 15.18& 0.99783 \\
1803+784 &  B &  0.684& 3.9 & 3.56  & 10.55&  6.04 &0.9862  \\
1823+568 &  B & 0.664  & 2.56 &  17.35 &3.29 &2.79 &0.93356\\
2007+776 &  B &  0.342 & 2.3& 6.73 &5.89 &  3.48 &  0.95782 \\
2200+420 &  B &0.069  & 3.7  & 10.24& 5.32 &  4.04 &0.96888 \\
1323+321 &  G  & 0.729& 1.7& 59 & 0.8   &1.73&     0.81601 \\
0108+388 &  G &0.669&  2.14 &40.66 &  1.18 &  2.95& 0.94079\\
0710+439 &  G &0.518 &1.25  &68.73& 0.64&  2.32 &  0.90234 \\
1845+797 &  G &0.0546 &1.82 &44.83& 1.16 & 2.44 &   0.91216\\
2021+614 &  G &  0.227 &0.2 &14.42  & 1.59 & 1.12& 0.45034 \\
0430+052  & G &  0.033 & 4 &   8.65 &6.13 & 4.45 & 0.97442 \\
0316+413  & G  &0.017  & 0.43 &  50.54& 0.33  & 1.97 &0.86158 \\
0415+379  & G &  0.049& 3.42 &  30.12& 1.05 & 6.6 & 0.98845\\
1830+285  & Q &  0.594 & 2.55  & 38.36  & 0.97 &4.35 &0.97322\\
2223-052  & Q & 1.404& 3.4 &0.65 & 24.24 & 12.38  & 0.99673\\
1721+343  &  Q& 0.206& 2.3& 46.64 & 0.23 &13.6 &0.99729 \\
1253-055&Q& 0.538& 9.2 &1.99& 21.09& 12.58 &0.99684\\
1226+023& Q &0.158  & 8& 8.48 &6.66& 8.21& 0.99255\\
2230+114& Q &1.037& 14.2 &7.85& 2.3 &45.13 &0.99975 \\
1222+216& Q&0.435 &1.4& 21.49& 2.48& 1.84& 0.83942\\
1828+487 &Q&0.691 &  8.32& 2.9& 16.14 &10.25& 0.99523 \\
1458+718&Q& 0.905 &6.53 &  4.79 &10.68& 7.38& 0.99078 \\
1633+382&Q& 1.814& 4.8 &5.48& 8.83& 5.78& 0.98492 \\
1641+399& Q&0.595& 9.5& 8.32& 6.37 &10.35& 0.99532 \\
1618+177 &Q&0.555 &1.88& 53.9& 0.46 &5.21& 0.98141\\
1928+738 &Q& 0.302& 7& 10.29& 5.4& 7.33& 0.99065 \\
1611+343& Q&1.401 & 11.4 &8.4& 5.04& 15.52& 0.99792 \\
1510-089*& Q&0.361 &3.77& 2.31 &13.18& 7.17 &0.99023 \\
1606+106*&Q&1.226 &2.9& 3.52 &9.32& 5.16 &0.98104 \\
1156+295& Q&0.729 &26.1 &4.03 &7.85 &47.38 &0.99978 \\
1150+812& Q&1.25 &4.1& 20.77& 2.41 &4.9& 0.97895 \\
1137+660& Q&0.652 &  1.3& 65.29& 0.71& 1.45& 0.72414 \\
0420-00& Q&0.844& 6.1& 4.17 &11.46& 7.4& 0.99083 \\
0420-014*&Q&0.915& 4.8 &3.45 &11.72& 6.89& 0.98941\\
0458-020*& Q &2.286 &4.09 &1.41& 17.8 &9.4& 0.99433 \\
0528+134&Q&2.07& 5.15& 2.59& 14.22 &8.08& 0.99231 \\
0538+498 &Q&0.545 &1.3& 0.58 &15.94 &8.05 &0.99225 \\
0336-01& Q &0.852&   8.9 &2.32 &19.01& 11.61& 0.99628 \\
0333+321& Q&1.258& 4.77& 1.37& 19.45 &  10.34& 0.99531 \\
0106+013&Q &2.107 &8.2& 1.54& 23.34& 13.34& 0.99719 \\
0153+744 &Q&2.34& 3.43& 20.41& 2.77& 3.69& 0.96258 \\
0202+149& Q&0.833 &0.4& 1.34& 5.93 &3.06 &0.94509 \\
0212+735& Q&2.37& 3.9 &3.07& 11.46 &6.44 &0.98787 \\
0234+285& Q&1.213& 9.29& 2.19 &19.99& 12.18& 0.99662 \\
0552+398& Q&2.365& 1.8 &6.02& 5.65& 3.2 &0.94992 \\
0605-08*& Q&0.872 &4.4& 12.75 &4.53& 4.51& 0.97511 \\
0923+392 &Q&0.699 &3.5 &1.83& 14.42 &7.67& 0.99146 \\
0016+731 &Q&1.781& 8.3 &4.03& 12.95& 9.17 &0.99404 \\
1039+811 &Q &1.26& 2.2 &40.99& 1.11& 3.19& 0.94959\\
1040+123& Q &1.029& 3.1 &24.76& 2.2& 3.51 &0.95856 \\
1055+018*& Q&0.888& 2.3& 4.06 &   7.78& 4.29& 0.97245 \\
0906+430&Q& 0.67& 3.86& 0.17 &51.64& 25.97& 0.99926 \\
0850+581& Q&1.332& 3.9& 14.63& 3.9 &4.03 &0.96872 \\
0615+820& Q&0.71& 2.2& 24.3& 2.43& 2.42& 0.91063 \\
0711+356 &Q&1.62& 5.5& 2.36& 15.41& 5.54 &0.98357 \\
0723+679 &Q&0.846 & 4.8 &23.3 &0.5& 24.35 &0.99916 \\
0827+243&Q&2.046 &12.08 &3.6 &15.46 &12.48 &0.99678 \\
0836+710& Q&2.17 &10.4& 5.51& 10.41 &10.45 &0.99541 \\
2251+158 &Q&0.859 &8.8 &7.83& 7.18& 9.05 &0.99388\\
\hline                                   %inserts single line
\end{tabular}
\end{table*}
}
\section{Results}

In this section, the main results related to superluminal motion are
displayed for different kinds of cosmological sources. I pay special
attention to the comparison between AGNs and GRBs in both radio and
$\gamma$ energy bands.

\subsection{Relations of $\beta_{app}$ with $\delta$, $z$ and $\gamma$}

\begin{figure}[h]
   \centering
 \includegraphics[width=12cm]{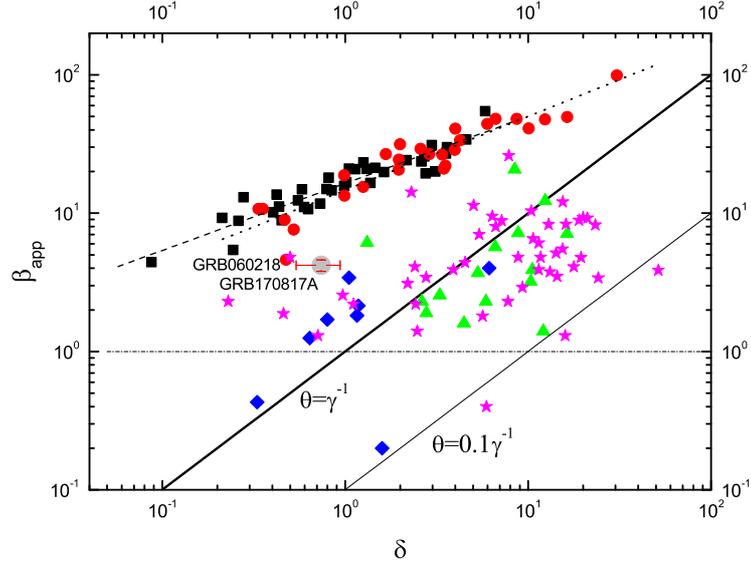}
      \caption{The apparent transverse velocity $\beta_{app}$ vs. the Doppler factor $\delta$ for AGNs and GRBs in our samples.
      \textit{Circles:}Swift GRBs; \textit{squares:}pre-Swift GRBs; \textit{stars:}quasars; \textit{triangles:}BL Lac objects;
      \textit{diamonds:} Radio galaxies. Thick, thin and dash-dotted lines denote the relations of $\beta_{app}=\delta$, $\beta_{app}=0.1\delta$
      and $\beta_{app}=constant$, respectively. The dotted and dashed lines correspond to their best fits to the GRB data.
      All the symbols defined here are same to those in the Figs. \ref{beta-redshift}, \ref{beta-gamma} and \ref{gamma-theta}.}
         \label{beta-delta}
   \end{figure}

Fig. \ref{beta-delta} shows a largely different population between
GRBs and AGNs in the plane of $\beta_{app}$ versus $\delta$. For
pre-Swift and Swift long GRBs, however, they exhibit a similar
positively correlated tendency. The dotted and dashed lines
correspond to their best fits to the GRB data, following a very
close power-law of $\beta_{app}\sim\delta^{0.5}$. Unlike GRBs, most
AGNs locate within the range of $\theta\geq0.1\gamma^{-1}$ and
roughly distribute along the line of $\beta_{app}\propto\delta$ with
larger dispersion. A linear correlative analysis gives the
correlation coefficient of 0.6 for all AGNs with a tiny chance
probability of $P<0.0001$.

\begin{figure}
   \centering
  \includegraphics[width=12cm]{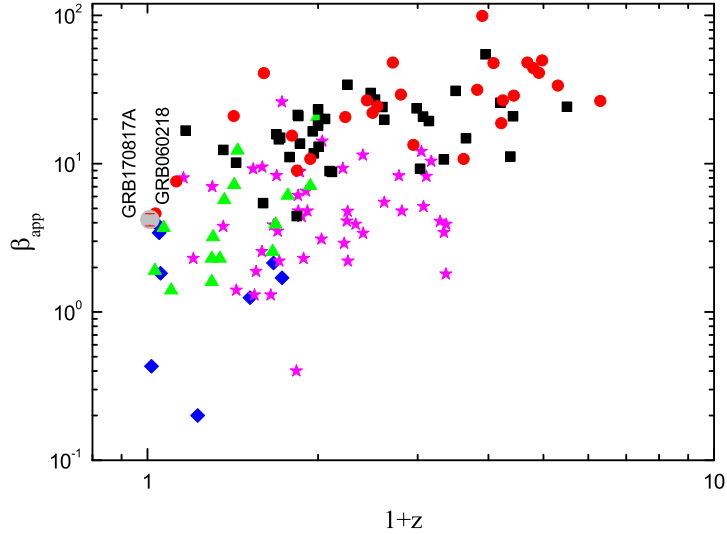}
      \caption{The apparent transverse velocity $\beta_{app}$ vs. cosmological inflation factor of $1+z$
       in the logarithmic scale.}
         \label{beta-redshift}
   \end{figure}

Another interesting phenomenon seen from Fig.\ref{beta-redshift} is
that $\beta_{app}$ obviously evolve with redshift for any kinds of
these cosmological sources. Furthermore, Fig.\ref{beta-redshift}
shows the positive correlation of $\beta_{app}$ with redshift
becomes more tight when all AGNs and GRBs are combined to be a whole
sample. The positive correlations demonstrate farther the sources
observed larger the apparent transverse velocity. Based on the three
relations between $\theta$ and $1/\gamma$ in section 2, we see that
$\beta_{app}\sim\gamma$ and $\beta_{app}\sim2/\theta$ when
$\theta\sim1/\gamma$ and $\theta>1/\gamma$ for AGNs and GRBs,
respectively. $\gamma$ and $\beta_{app}$ are tightly correlated for
the different radio-loud AGNs, while there is no any relation for
LGRBs. This is because the viewing angle $\theta$ is far large than
$1/\gamma$, which cause $\beta_{app}$ becomes only dependent on
$\theta$. However, AGNs with $\theta\sim1/\gamma$ seen from
Fig.\ref{beta-delta} make them locate around the line of
$\beta_{app}\sim\gamma$.

\begin{figure*}
   \centering
  \includegraphics[width=18cm]{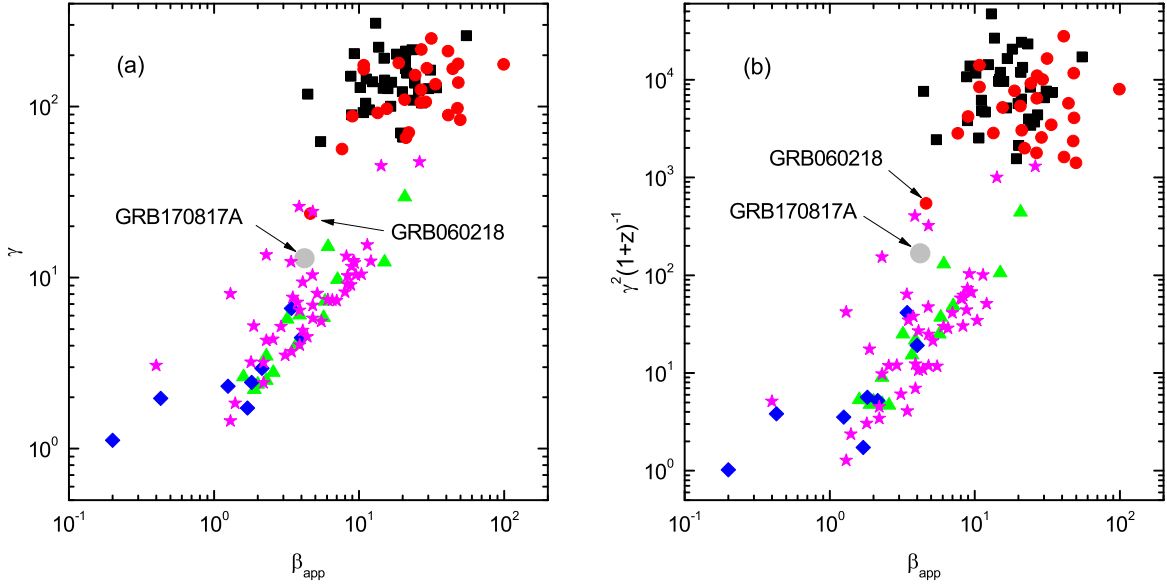}
      \caption{Correlations of the apparent velocity $\beta_{app}$ with the Lorentz factor $\gamma$ in panel (a) and the modificatory
      factor $\gamma^2(1+z)^{-1}$ for the effects of Doppler boosting and cosmological expansion in panel (b), respectively. }
         \label{beta-gamma}
   \end{figure*}

Plot of $\beta_{app}$ vs. $\gamma$ in Fig.\ref{beta-gamma}a shows
how the apparent transverse velocity is affected by Lorentz factor
for all the radio-loud AGNs and LGRBs. It is interestingly found
that the $\beta_{app}$ is positively correlated with $\gamma$, while
this trend disappears for GRBs. Fig.\ref{beta-gamma}b indicates that
the intrinsic observations, e.g. some typically comoving timescale
in source frame, reduced for cosmological inflation and Doppler
effect still hold above similar properties for different AGNs and
GRBs. This means AGNs and GRBs are basically distinct cosmological
sources in nature.

In addition, one can find in Figs. 2, 3 and 4 that radio galaxies are
distributed in the small end of $\beta_{app}$, largely different
from BL Lac objects and Quasars, much less the LGRBs. This means the
radio galaxies with lower radio luminosity are very special and may
thus result from a distinct origin, compared with other radio-loud
AGNs such as BL Lac objects and Quasars.

\subsection{Correlation between $\gamma$ and $\theta$}

\begin{figure}
   \centering
  \includegraphics[width=12cm]{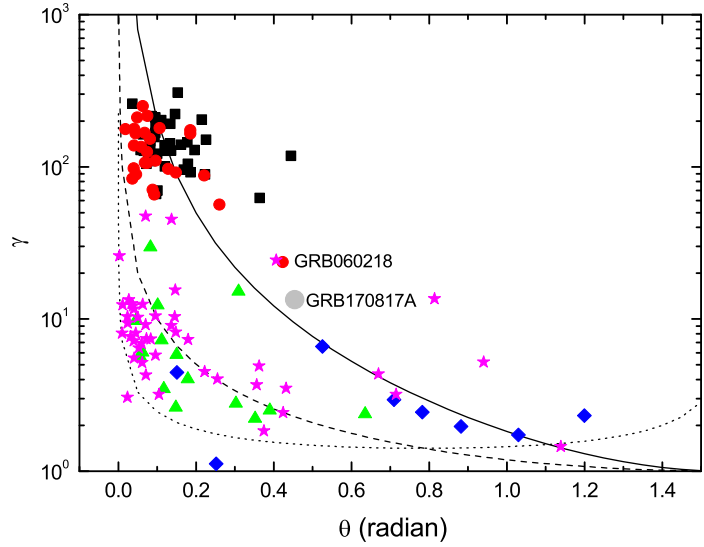}
      \caption{The Lorentz factor $\gamma$ vs. the viewing angle $\theta$. \textit{Solid line}
   corresponds to $\delta=1$; \textit{dashed line} represents $\gamma=1/sin\theta$; and \textit{dotted
   line} describes $\beta_{app}=1$.}
         \label{gamma-theta}
   \end{figure}

In Fig.\ref{gamma-theta}, we compare these sources in the $\gamma$
vs. $\theta$ plane. The top-right region of $\delta=1$ corresponds
to $\delta<1$ and the reverse direction is within the range of
$\delta>1$. Similarly, the bottom-left of $\beta_{app}=1$ represents
$\beta_{app}<1$ and the top-right of $\beta_{app}=1$ stands for
$\beta_{app}>1$. Very clearly, most of AGNs and GRBs reside between
the solid line and the dotted one, in the area of $\beta_{app}>1$
and $\delta>1$. The result on AGNs is extremely consistent with that
suggested by Ghisellini (1993) in theory. But we again find the
radio galaxies with lower lorentz factor compared to other AGNs
locate at the larger end of $\theta$. Another interesting phenomenon
is GRBs with smaller viewing angle distribute at the larger end of
$\gamma$. This indicates that the beaming effect could be really not
very important for the radio galaxies, unlike for GRBs, BL Lac
objects and Quasars. A majority of AGNs follows
$\gamma=1/\sin\theta$, that is $\beta_{app}=\beta\delta$, while GRBs
cluster along the line of $\delta=1$. In fact, the locations of GRBs
might move towards the short end of $\theta$ slightly if the jet
evolution is considered.

\subsection{Violators}
It is well-known that GRB 060218 is a typical low luminosity burst ($T_{90}\sim2100 s$) connected with SN 2006aj, a supernovae Ib/c (e.g. Mirabal et al. 2006; Li 2007). Recently, Zhang et al. (2018a) noticed that GRB 060218, together with 980425 and 031203, deviated from the normal evolution curves of long bursts in the plane of peak flux density versus redshift at different radio frequencies. GRB 170817A is a peculiarly newly-discovered short GRB ($T_{90}\sim2 s$) associated with gravitational wave (GW), which was reported as the first electromagnetic counterpart of a GW event originated from binary Neutron stars (Abbott et al. 2016a,b). With more and more short bursts accumulated with redshift and peak energy, their spectrum-energy relations can be built as long bursts did before. Excitingly, three spectrum-energy correlations have been proposed by Zhang et al. (2018b) and the GRB 170817A, unlike other short and long bursts, is identified as an outlier to all the three energy relations of short GRBs. It is interesting to note here that GRB 170817A is also found to be very different from long bursts but similar to GRB 060218 regarding the superluminal motion. This may hint that those low luminosity GRBs could hold the consistent observations related with the superluminal motions.

\section{Conclusions and discussions}

Our main results are summarized and concluded in the following:

\begin{enumerate}
\item The apparent transverse velocity $\beta_{app}$ is found to be
correlated with $\delta$ as $\beta_{app}\propto\delta$ for LGRBs and
$\beta_{app}\propto\delta^{0.5}$ for radio-loud AGNs.
\item Radio galaxies are special AGNs with lower
redshift, luminosity, $\gamma$ and $\delta$, which makes it a very
different radio-loud sources.
\item Figs. \ref{beta-delta},\ref{beta-gamma} and \ref{gamma-theta} seem to demonstrate GRBs and AGNs may have
different radiation mechanisms and Doppler (or beaming)effect.
However, their cosmological evolution might be same as shown in Fig.
\ref{beta-redshift} in respect of the superluminal motion.
\item As low luminosity GRBs, 060218 and 170817A are obvious violators to the above relation of both long GRBs and three types of AGNs and
have very similar behaviors although they are largely different in duration time $T_{90}$.
\end{enumerate}

As suggested by Kellermann et al. (2004), radio jets that are also
strong gamma-ray sources detected by EGRET appear to have
significantly faster speeds than non-EGRET sources, which is much
consistent with the idea that gamma-ray sources have larger Doppler
factors than non¨Cgamma-ray sources. Kellermann et al. (2004) also
pointed that there is a systematic decrease in apparent velocity
with increasing wavelength, probably because the observations at
different wavelengths sample built from different parts of the jet
structure. It is required to test whether this is also true for GRBs
because the current sample with with superluminal motion for GRB
radio afterglows are very limited, ONLY GRB 030329 and possibly GRB
980425. In such a case, observational investigations on superluminal
motion of GRBs are particularly expected in the near future.

The typically extragalactic sources with high redshift are usually
useful for studying the formation and evolution of the early
universe. So far, the farthest galaxy is found to have a
spectroscopic redshift of z = 6.96 (Iye et al. 2006 ). Fan et al.
(2001) found the most distant AGN was Quasar SDSS J1030+0524 with a
redshift of $z = 6.28$, which was taken among follow-up spectroscopy
of i-band drop-out objects. Cosmological GRBs are the ideal
candidates for the universal study since several GRBs have been
found to locate at very high universe with redshifts larger than
$z\sim6$, e.g., GRB 080913 ($z = 6.7$, Greiner, J. et al. 2009), GRB
050904 (z = 6.39, Haislip et al. 2006) and GRB090423 ($z = 8.1$,
Salvaterra et al. 2009). In principle, superluminal motions should
be detectable and similar for all the distant celestial bodies if
only the true velocity of the outflow is close to the speed of light
and the viewing angle $\theta$ is sufficiently small, due to the
beaming effect of special relativity (see also Zhang et al. 2007).
In addition, this phenomenon would be multiple-band observed if the
receiver on ground is sensitive enough. However, studies on
superluminal motions of these sources proceed slowly till now owing
to the lack of radio observations with high resolution.

In statistics, roughly 15-20\% of AGNs are radio-loud. The
variability of radio-loud AGNs behaves much active over wide
wavelengths (Xie et al. 1994). Unlike the steep-spectrum for
radio-quiet AGNs, the spectrum of radio-loud AGNs is generally flat
because of the more beamed outflows at radio frequencies, but not in
the optical and X-ray energy bands. This might imply that radio-loud
AGNs are jet-dominated at radio frequencies and disk-dominated in
the X-ray and optical band. Even though, the radio mechanism of the
radio-loud AGNs could be very different from that of LGRBs. The
later with a single or double power-law form is thought to be
basically produced from the processes of synchrotron radiation and
inverse-Compton scattering. Currently, the real radiation mechanism
of AGNs and GRBs are still not confirmed and needs to be further
clarified.

\begin{acknowledgements}
This work is supported by the National Natural Science Foundation of
China (grant number: U1431126, 11263002), Guizhou natural and scientific
fundings (grant numbers: 20165660, 201519, OP201511 and 114A11KYSB20160008).

\end{acknowledgements}

\end{document}